# Pruning Attribute Values From Data Cubes with Diamond Dicing

Hazel Webb and Owen Kaser
University of New Brunswick
hazel.webb@unb.ca, o.kaser@computer.org

Daniel Lemire
Université du Québec à Montréal
lemire@acm.org

October 28, 2018

**Abstract**

Data stored in a data warehouse are inherently multidimensional, but most data-pruning techniques (such as iceberg and top-$k$ queries) are unidimensional. However, analysts need to issue multidimensional queries. For example, an analyst may need to select not just the most profitable stores or—separately—the most profitable products, but simultaneous sets of stores and products fulfilling some profitability constraints. To fill this need, we propose a new operator, the *diamond dice*. Because of the interaction between dimensions, the computation of diamonds is challenging.

We present the first diamond-dicing experiments on large data sets.

Experiments show that we can compute diamond cubes over fact tables containing 100 million facts in less than 35 minutes using a standard PC.

## 1 Introduction

In signal and image processing, software subsamples data [29] for visualization, compression, or analysis purposes: commonly, images are cropped to focus the attention on a segment. In databases, researchers have proposed similar subsampling techniques [3,14], including iceberg queries [13,27,33] and top-k queries [21, 22]. Formally, subsampling is the selection of a subset of the data, often with desirable properties such as representativity, conciseness, or homogeneity. Of the subsampling techniques applicable to OLAP, only the dice operator focuses on reducing the number of attribute values without aggregation whilst retaining the original number of dimensions.

Such reduced representations are sometimes of critical importance to get good online performance in Business Intelligence (BI) applications [2, 13]. Even when performance is not an issue, browsing and visualizing the data frequently benefit from reduced views [4].

Often, business analysts are interested in distinguishing elements that are most crucial to their business, such as the $k$ products jointly responsible for 50% of all sales, from the *long tail* [1]—the lesser elements. The computation of icebergs, top-k elements, or heavy-hitters has received much attention [7–9]. We wish to generalize this type of query so that interactions between dimensions are allowed. For example, a business analysts might want to compute a small set of stores and business hours jointly responsible for over 80% of the sales. In this new setting, the head and tails of the distributions must be described using a multidimensional language; computationally, the queries become significantly more difficult. Hence, analysts will often process dimensions one at a time: perhaps they would focus first on the most profitable business hours, and then aggregate sales per store, or perhaps they would find the must profitable stores and aggregate sales per hour. We propose a general model, of which the unidimensional analysis is a special case, that has acceptable computational costs and a theoretical foundation. In the two-dimensional case, our proposal is a generalization of ITERATIVE PRUNING [18], a graph-trawling approach used to analyze social networks. It also generalizes iceberg queries [13, 27, 33].

Table 1: Sales (in million dollars) with a 4,10 sum-diamond shaded: stores need to have sales above $10 million whereas product lines need sales above $4 million

|  | Chicago | Montreal | Miami | Paris | Berlin |
|---|---|---|---|---|---|
| TV | 3.4 | 0.9 | 0.1 | 0.9 | 2.0 |
| Camcorder | 0.1 | 1.4 | 3.1 | 2.3 | 2.1 |
| Phone | 0.2 | 8.4 | 2.1 | 4.5 | 0.1 |
| Camera | 0.4 | 2.7 | 6.3 | 4.6 | 3.5 |
| Game console | 3.2 | 0.3 | 0.3 | 2.1 | 1.5 |
| DVD Player | 0.2 | 0.5 | 0.5 | 2.2 | 2.3 |

To illustrate our proposal in the BI context, consider the following example. Table 1 represents the sales of different items in different locations. Typical iceberg queries might be requests for stores having sales of at least 10 million dollars or product lines with sales of at least 4 million dollars. However, what if the analyst wants to apply both thresholds simultaneously? He might contemplate closing both some stores and some product lines. In our example, applying the constraint on stores would close Chicago, whereas applying the constraint on product lines would not terminate any product line. However, once the shop in Chicago is closed, we see that the product line TV must be terminated which causes the closure of the Berlin store and the termination of two new product lines (Game console and DVD player).

This multidimensional pruning query selects a subset of attribute values from each dimension that are simultaneously important. The operation is a *diamond dice* [32] and produces a *diamond*, as formally defined in Section 3.

Other approaches that seek important attribute values, e.g. the Skyline operator [6, 23], Dominant Relationship Analysis [20], and Top-$k$ dominating queries [35], require dimension attribute values to be ordered, e.g. distance between a hotel and a conference venue, so that data points can be ordered. Our approach requires no such ordering.

## 2  Notation

Notation used in this paper is tabulated below.

| | |
|---|---|
| $C$ | a data cube |
| $\sigma$ | aggregator COUNT or SUM |
| $C_{\text{dim},j}$ | $\sigma$(slice $j$ of dimension dim in cube $C$) |
| $\|C\| = \sum_j C_{1,j}$ | the number of allocated cells in cube $C$ |
| $A$, $B$ | cubes |
| $D_i$ | $i^{\text{th}}$ dimension of a data cube |
| $n_i$ | number of attribute values in dimension $D_i$ |
| $k$ | number of carats |
| $k_i$ | number of carats of order 1 for $D_i$ |
| $d$ | number of dimensions |
| $p$ | max. number of attribute values per dim |
| $p_i$ | max. number of attribute values for $D_i$ |
| $\kappa(C)$ | maximum carats in $C$ |
| COUNT-$\kappa(C)$ | maximum carats in $C$, $\sigma$ is COUNT |

## 3 Properties of Diamond Cubes

Given a database relation, a dimension $D$ is the set of values associated with a single attribute. A cube $C$ is the set of dimensions together with a map from some tuples in $D_1 \times \cdots \times D_d$ to real-valued measure values. Without losing generality, we shall assume that $n_1 \leq n_2 \leq \ldots \leq n_d$, where $n_i$ is the number of distinct attribute values in dimension $i$.

A *slice* of order $\delta$ is the set of cells we obtain when we fix a single attribute value in each of $\delta$ different dimensions. For example, a slice of order 0 is the entire cube, a slice of order 1 is the more traditional definition of a slice and so on. For a $d$-dimensional cube, a slice of order $d$ is a single cell. An aggregator is a function, $\sigma$, from sets of values to the real numbers.

**Definition 1.** *Let $\sigma$ be an aggregator such as* SUM *or* COUNT*, and let $k$ be some real-valued number. A cube has $k$ carats over dimensions $i_1, \ldots, i_\delta$, if for every slice $x$ of order $\delta$ along dimensions $i_1, \ldots, i_\delta$, we have $\sigma(x) \geq k$.*

We can recover iceberg cubes by seeking cubes having carats of order $d$ where $\sigma(x)$ returns the measure corresponding to cell $x$. The predicate $\sigma(x) \leq k$ could be generalized to include $\sigma(x) \geq k$ and other constraints.

We say that an aggregator $\sigma$ is monotonically increasing if $S' \subset S$ implies $\sigma(S') \leq \sigma(S)$. Similarly, $\sigma$ is monotonically decreasing if $S' \subset S$ implies $\sigma(S') \geq \sigma(S)$. Monotonically increasing operators include COUNT, MAX and SUM (over non-negative measures). Monotonically decreasing operators include MIN and SUM (over non-positive measures).

We say a cube $C'$ is *restricted from* cube $C$ if

- they have the same number of dimensions

- dimension $i$ of $C'$ is a subset of dimension $i$ of $C$

- If in $C'$, $(v_1, v_2, \ldots, v_d) \mapsto m$, then in $C$, $(v_1, v_2, \ldots, v_d) \mapsto m$

**Definition 2.** *Let $A$ and $B$ be two cubes with the same dimensions and measures restricted from a single cube $C$. Their union is denoted $A \cup B$. It is the set of attributes together with their measures, on each dimension, that appear in $A$, or $B$ or both. The union of $A$ and $B$ is $B$ if and only if $A$ is contained in $B$: $A$ is a subcube of $B$.*

**Proposition 1.** *If the aggregator $\sigma$ is monotonically increasing, then the union of any two cubes having $k$ carats over dimensions $i_1, \ldots, i_\delta$ has $k$ carats over dimensions $i_1, \ldots, i_\delta$ as well.*

*Proof.* Any slice $x$ of the union of $A$ and $B$ contains a slice $x'$ from at least $A$ or $B$. Since $x'$ is contained in $x$, and $\sigma(x') \geq k$, we have $\sigma(x) \geq k$. □

Hence, as long as $\sigma$ is monotonically increasing, there is a maximal cube having $k$ carats over dimensions $i_1, \ldots, i_\delta$, and we call such a cube the *diamond*. When $\sigma$ is not monotonically increasing, there may not be a unique diamond. Indeed, consider the even-numbered rows and columns of the following matrix, then consider the odd-numbered rows and columns. Both are maximal cubes with 2 carats (of order 1) under the SUM operator:

$$\begin{array}{rrrr} 1 & -1 & 1 & -1 \\ -1 & 1 & -1 & 1 \\ 1 & -1 & 1 & -1 \\ -1 & 1 & -1 & 1 \end{array}$$

Because we wish diamonds to be unique, we will require $\sigma$ to be

The next proposition shows that diamonds are themselves nested.

**Proposition 2.** *The diamond having $k'$ carats over dimensions $i_1, \ldots, i_\delta$ is contained in the diamond having $k$ carats over dimensions $i_1, \ldots, i_\delta$ whenever $k' \geq k$.*

*Proof.* Let $A$ be the diamond having $k$ carats and $B$ be the diamond having $k'$ carats. By Proposition 1, $A \cup B$ has at least $k'$ carats, and because $B$ is maximal, $A \cup B = B$; thus, $A$ is contained in $B$. □

For simplicity, we only consider carats of order 1 for the rest of the paper. We write that a cube has $k_1, k_2, \ldots, k_d$-carats if it has $k_i$ carats over dimension $D_i$; when $k_1 = k_2 = \ldots = k_d = k$ we simply write that it has $k$ carats.

One consequence of Proposition 2 is that the diamonds having various number of carats form a lattice (see Fig. 1) under the relation "is included in." This lattice creates optimization opportunities: if we are given the $2, 1$-carat diamond $X$ and the $1, 2$-carat diamond $Y$, then we know that the $2, 2$-carat diamond must lie in both $X$ and $Y$. Likewise, if we have the $2, 2$-carat diamond, then we know that its attribute values must be included in the diamond above it in the lattice (such as the $2, 1$-carat diamond).

Given the size of a sum-based diamond cube (in cells), there is no upper bound on its number of carats. However, it cannot have more carats than the sum of its measures. Conversely, if a cube has dimension sizes $n_1, n_2, \ldots, n_d$ and $k$ carats, then its sum is at least $k \max(n_1, n_2, \ldots, n_d)$.

Given the dimensions of a COUNT-based diamond cube, $n_1 \leq n_2 \leq \ldots \leq n_{d-1} \leq n_d$, an upper bound for the number of carats $k$ of a subcube is $\prod_{i=1}^{d-1} n_i$. An upper bound on the number of carats $k_i$ for dimension $i$ is $\prod_{j=1, j \neq i}^{d} n_i$. An alternate (and trivial) upper bound on the number of carats in any dimension is $|C|$, the number of allocated cells in the cube. For sparse cubes, this bound may be more useful.

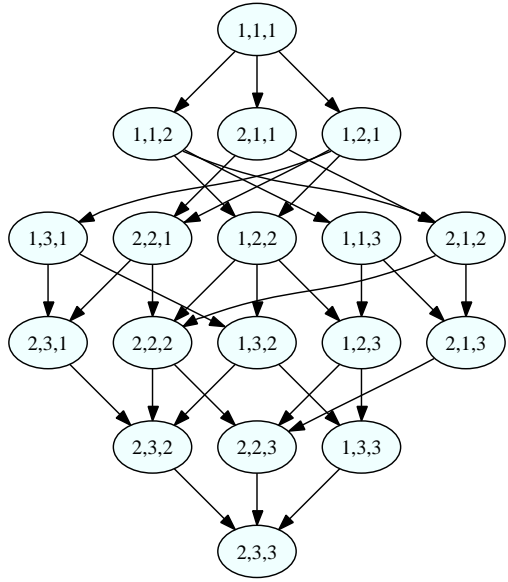

Figure 1: Part of the COUNT-based diamond-cube lattice of a $2 \times 2 \times 2$ cube

Intuitively, a cube with many carats needs to have a large number of allocated cells: accordingly, the next proposition provides a lower bound on the size of the cube given the number of carats.

**Proposition 3.** *For $d > 1$, the size $S$, or number of allocated cells, of a d-dimensional cube of $k$ carats satisfies $S \geq k \max_{i \in \{1,2,\ldots,d\}} n_i \geq k^{d/(d-1)}$; more generally, a $k_1, k_2, \ldots, k_d$-carat cube has size $S$ satisfying $S \geq \max_{i \in \{1,2,\ldots,d\}} k_i n_i \geq (\prod_{i=1,\ldots,d} k_i)^{1/(d-1)}$.*

*Proof.* Pick dimension $D_i$: the subcube has $n_i$ slices along this dimension, each with $k$ allocated cells, proving the first item.

We have that $k(\sum_i n_i)/d \leq k \max_{i \in \{1,2,\ldots,d\}} n_i$ so that the size of the subcube is at least $k(\sum_i n_i)/d$.

If we prove that $\sum_i n_i \geq d k^{1/(d-1)}$ then we will have that $k(\sum_i n_i)/d \geq k^{d/(d-1)}$ proving the second item. This result can be shown using Lagrange multipliers. Consider the problem of minimizing $\sum_i n_i$ given the constraints $\prod_{i=1,2,\ldots,j-1,j+1,\ldots,d} n_i \geq k$ for $j = 1, 2, \ldots, d$. These constraints are necessary since all slices must contain at least $k$ cells. The corresponding Lagrangian is $L = \sum_i n_i + \sum_j \lambda_j (\prod_{i=1,2,\ldots,j-1,j+1,\ldots,d} n_i - k)$. By inspection, the derivatives of $L$ with respect to $n_1, n_2, \ldots, n_d$ are zero and all constraints are satisfied when $n_1 = n_2 = \ldots = n_d = k^{1/(d-1)}$. For these values, $\sum_i n_i = d k^{1/(d-1)}$ and this must be a minimum, proving the result. The more general result follows similarly, by proving that the minimum of $\sum n_i k_i$ is reached when $n_i = (\prod_{i=1,\ldots,d} k_i)^{1/(d-1)}/k_i$ for all $i$'s. □

We calculate the *volume* of a cube $C$ as $\prod_{i=1}^{i=d} n_i$ and its *density* is the ratio of allocated cells, $|C|$, to the volume ($|C|/\prod_{i=1}^{i=d} n_i$). Given $\sigma$, its *carat-number*, $\kappa(C)$, is the largest number of carats for which the cube has a non-empty diamond. Intuitively, a small cube with many allocated cells should have a large $\kappa(C)$.

One statistic of a cube $C$ is its carat-number, $\kappa(C)$, which is the largest number of carats for which the cube has a non-empty diamond. Is this statistic robust? I.e., with high probability, can changing a small fraction of the data set change the statistic much? Of course, typical analyses are based on thresholds (e.g. applied to support and accuracy in rule mining), and thus small changes to the cube may not always behave as desired. Diamond dicing is no exception. For the cube $C$ in Fig. 3 and the statistic $\kappa(C)$ we see that diamond dicing is not robust against an adversary who can deallocate a single cell: deallocation of the second cell on the top row results means that the cube no longer contains a diamond with 2 carats. This example can be generalized.

**Proposition 4.** *For any $b$, there is a cube $C$ from which deallocation of any $b$ cells results in a cube $C'$ with $\kappa(C') = \kappa(C) - \Omega(b)$.*

*Proof.* Let $C$ be a $d$-dimensional cube with $n_i = 2$ with all cells allocated. We see that $C$ has $2^{d-1}$ carats and $\kappa(C) = 2^{d-1}$ (assume $d > 1$). Given $b$, set $x = \lfloor \frac{(d-1)b}{2d} \rfloor$. Because $x \geq \frac{(d-1)b}{2d} - 1 \geq \frac{b}{4} - 1 \in \Omega(b)$, it suffices to show that by deallocating $b$ cells, we can reduce the number of carats by $x$. By Proposition 3, we have that any cube with $2^{d-1} - x$ carats must have size at least $(2^{d-1} - x)^{d/(d-1)}$. When $x \ll 2^{d-1}$, this size is approximately $2^{d-1} - \frac{2dx}{d-1}$, and slightly larger by the alternation of the Taylor expansion. Hence, if we deallocate at least $\frac{2dx}{d-1}$ cells, the number of carats must go down by at least $x$. But $x = \lfloor \frac{(d-1)b}{2d} \rfloor \Rightarrow x \leq \frac{(d-1)b}{2d} \Rightarrow b \geq \frac{2dx}{d-1}$ which shows the result. It is always possible to choose $d$ large enough so that $x \ll 2^{d-1}$ irrespective of the value $b$. □

Conversely, in Fig. 3 we might allocate the cell above the bottom-right corner, thereby obtaining a 2-carat diamond with all $2n+1$ cells. Compared to the original case with a 4-cell 2-carat diamond, we see that a small change effects a very different result. Diamond dicing is not, in general, robust. However, it is perhaps more reasonable to follow Pensa and Boulicaut [28] and ask whether $\kappa$ appears, experimentally, to be robust against random noise on realistic data sets. We return to this in Subsection 6.5.

Many OLAP aggregators are distributive, algebraic and linear. An aggregator $\sigma$ is *distributive* [16] if there is a function $F$ such that for all $0 \leq k < n-1$,

$$\sigma(a_0, \ldots, a_k, a_{k+1}, \ldots, a_{n-1}) = F(\sigma(a_0, \ldots, a_k), \sigma(a_{k+1}, \ldots, a_{n-1})).$$

An aggregator $\sigma$ is *algebraic* if there is an intermediate tuple-valued **distributive** range-query function $G$ from which $\sigma$ can be computed. An algebraic example is AVERAGE: given the tuple (COUNT, SUM), one can compute AVERAGE by a ratio. In other words, if $\sigma$ is an algebraic function then there must exist $G$ and $F$ such that

$$G(a_0, \ldots, a_k, a_{k+1}, \ldots, a_{n-1}) = F(G(a_0, \ldots, a_k), G(a_{k+1}, \ldots, a_{n-1})).$$

An algebraic aggregator $\sigma$ is *linear* [19] if the corresponding intermediate query $G$ satisfies

$$G(a_0 + \alpha d_0, \ldots, a_{n-1} + \alpha d_{n-1}) = G(a_0, \ldots, a_{n-1}) + \alpha G(d_0, \ldots, d_{n-1})$$

for all arrays $a, d$, and constants $\alpha$. SUM and COUNT are linear functions; MAX is not linear.

## 4  Related Problems

In this section, we discuss four problems, three of which are NP-hard, and show that the diamond—while perhaps not providing an exact solution—is a good starting point. The first two problems, Trawling the Web for Cyber-communities and Largest Perfect Subcube, assume use of the aggregator COUNT whilst for the remaining problems we assume SUM.

### 4.1  Trawling the Web for Cyber-communities

In 1999, Kumar et al. [18] introduced the ITERATIVE PRUNING algorithm for discovering emerging communities on the Web. They model the Web as a directed graph and seek large dense bipartite subgraphs or cores, and therefore their problem is a 2-D version of our problem. Although their paper has been widely cited [30, 34], to our knowledge, we are the first to propose a multidimensional extension to their problem suitable for use in more than two dimensions and to provide a formal analysis.

### 4.2  Largest Perfect Cube

A *perfect* cube contains no empty cells, and thus it is a diamond. Finding the largest perfect diamond is NP-hard. A motivation for this problem is found in Formal Concept Analysis [15], for example.

**Proposition 5.** *Finding a perfect subcube with largest volume is NP-hard, even in 2-D.*

*Proof.* A 2-D cube is essentially an unweighted bipartite graph. Thus, a perfect subcube corresponds directly to a *biclique*—a clique in a bipartite graph. Finding a biclique with the largest number of edges has been shown NP-hard by Peeters [26], and this problem is equivalent to finding a perfect subcube of maximum volume. □

Finding a diamond might be part of a sensible heuristic to solve this problem, as the next lemma suggests.

**Lemma 1.** *For COUNT-based carats, a perfect subcube of size $n_1 \times n_2 \times \ldots \times n_d$ is contained in the $\prod_{i=1}^{d} n_i / \max_i n_i$-carat diamond and in the $k_1, k_2, \ldots, k_d$-carat diamond where $k_i = \prod_{j=1}^{d} n_j / n_i$.*

This helps in two ways: if there is a nontrivial diamond of the specified size, we can search for the perfect subcube within it; however, if there is only an empty diamond of the specified size, there is no perfect subcube.

### 4.3  Densest Cube with Limited Dimensions

In the OLAP context, given a cube, a user may ask to "find the subcube with at most 100 attribute values per dimension." Meanwhile, he may want to keep as much of the cube as possible. We call this problem DENSEST CUBE WITH LIMITED DIMENSIONS (DCLD), which we formalize as: pick $\min(n_i, p)$ attribute values for dimension $D_i$, for all $i$'s, so that the resulting subcube is maximally dense.

Intuitively, a densest cube should at least contain a diamond. We proceed to show that a sufficiently dense cube always contains a diamond with a large number of carats.

**Proposition 6.** *If a cube does not contain a $k$-carat subcube, then it has at most $1 + (k-1)\sum_{i=1}^{d}(n_i - 1)$ allocated cells. Hence, it has density at most $(1 + (k-1)\sum_{i=1}^{d}(n_i - 1))/\prod_{i=1}^{d} n_i$. More generally, a cube that does not contain a $k_1, k_2, \ldots, k_d$-carat subcube has size at most $1 + \sum_{i=1}^{d}(k_i - 1)(n_i - 1)$ and density at most $(1 + \sum_{i=1}^{d}(k_i - 1)(n_i - 1))/\prod_{i=1}^{d} n_i$.*

*Proof.* Suppose that a cube of dimension at most $n_1 \times n_2 \times \ldots \times n_d$ contains no $k$-carat diamond. Then one slice must contain at most $k - 1$ allocated cells. Remove this slice. The amputated cube must not contain a $k$-carat diamond. Hence, it has one slice containing at most $k - 1$ allocated cells. Remove it. This iterative process can continue at most $\sum_i (n_i - 1)$ times before there is at most one allocated cell left: hence, there are at most $(k-1)\sum_i (n_i - 1) + 1$ allocated cells in total. The more general result follows similarly. □

The following corollary follows trivially from Proposition 6:

**Corollary 1.** *A cube of size greater than $1 + (k-1)\sum_{i=1}^{d}(n_i - 1)$ allocated cells, that is, having density greater than*
$$\frac{1 + (k-1)\sum_{i=1}^{d}(n_i - 1)}{\prod_{i=1}^{d} n_i},$$
*must contain a $k$-carat subcube. If a cube contains more than $1 + \sum_{i=1}^{d}(k_i - 1)(n_i - 1)$ allocated cells, it must contain a $k_1, k_2, \ldots, k_d$-carat subcube.*

Solving for $k$, we have a lower bound on the maximal number of carats: $\kappa(C) \geq |C|/\sum_i (n_i - 1) - 3$.

We also have the following corollary to Proposition 6:

**Corollary 2.** *Any solution of the DCLD problem having density above*
$$\frac{1 + (k-1)\sum_{i=1}^{d}(\min(n_i, p) - 1)}{\prod_{i=1}^{d} \min(n_i, p)} \leq \frac{1 + (k-1)d(p-1)}{\prod_{i=1}^{d} n_i}$$
*must intersect with the $k$-carat diamond.*

When $n_i \geq p$ for all $i$, then the density threshold of the previous corollary is $(1 + (k-1)d(p-1))/p^d$: this value goes to zero exponentially as the number of dimensions increases.

We might hope that when the dimensions of the diamond coincide with the required dimensions of the densest cube, we would have a solution to the DCLD problem. Alas, this is not true. Consider the 2-D cube in Fig. 2. The bottom-right quadrant forms the largest 3-carat subcube. In the bottom-right quadrant, there are 15 allocated cells whereas in the upper-left quadrant there are 16 allocated cells. This proves the next result.

**Lemma 2.** *Even if a diamond has exactly $\min(n_i, p_i)$ attribute values for dimension $D_i$, for all $i$'s, it may still not be a solution to the DCLD problem.*

We are interested in large data sets; the next theorem shows that solving DCLD and HCLD is difficult.

**Theorem 1.** *The DCLD and HCLD problems are NP-hard.*

Figure 2: Example showing that a diamond (bottom-right quadrant) may not have optimal density.

*Proof.* The EXACT BALANCED PRIME NODE CARDINALITY DECISION PROBLEM (EBPNCD) is NP-complete [10]—for a given bipartite graph $G = (V_1, V_2, E)$ and a number $p$, does there exist a biclique $U_1$ and $U_2$ in $G$ such that $|U_1| = p$ and $|U_2| = p$?

Given an EBPNCD instance, construct a 2-D cube where each value of the first dimension corresponds to a vertex of $V_1$, and each value of the second dimension corresponds to a vertex of $V_2$. Fill cell corresponding to $v_1, v_2 \in V_1 \times V_2$ with a measure value if and only if $v_1$ is connected to $v_2$. The solution of the DCLD problem applied to this cube with a limit of $p$ will be a biclique if such a biclique exists. □

It follows that HCLD is also NP-hard by reduction of DCLD.

### 4.4 Heaviest Cube with Limited Dimensions

In the OLAP context, given a cube, a user may ask to "find a subcube with 10 attribute values per dimension." Meanwhile, he may want the resulting subcube to have maximal average—he is, perhaps, looking for the 10 attributes from each dimension that, in combination, give the greatest profit. Note that this problem does not restrict the number of attribute values ($p$) to be the same for each dimension.

We call this problem the HEAVIEST CUBE WITH LIMITED DIMENSIONS (HCLD), which we formalize as: pick $\min(n_i, p_i)$ attribute values for dimension $D_i$, for all $i$'s, so that the resulting subcube has maximal average. We have that the HCLD must intersect with diamonds.

**Theorem 2.** *Using the SUM operator, a cube without any $k_1, k_2, \ldots, k_d$-carat subcube has sum less than $\sum_{i=1}^{d}(n_i + 1)k_i + \max(k_1, k2, \ldots, k_d)$ where the cube has size $n_1 \times n_2 \times \ldots \times n_d$.*

*Proof.* Suppose that a cube of dimension $n_1 \times n_2 \times \ldots \times n_d$ contains no $k_1, k_2, \ldots, k_d$-sum-carat cube. Such a cube must contain at least one slice with sum less than $k$, remove it. The remainder must also not contain a $k$-sum-carat cube, remove another slice and so on. This process may go on at most $\sum_{i=1}^{d}(n_i + 1)$ times before there is only one cell left. Hence, the sum of the cube is less than $\sum_{i=1}^{d}(n_i + 1)(k_i) + \max(k_1, k2, \ldots, k_d)$. □

**Corollary 3.** *Any solution to the HCLD problem having average greater than*

$$\frac{\sum_{i=1}^{d}(n_i + 1)k_i + \max(k_1, k2, \ldots, k_d)}{\prod_{i=1}^{d} n_i}$$

*must intersect with the $k_1, k_2, \ldots, k_d$-sum-carat diamond.*

## 5 Algorithm

We have developed and implemented an algorithm for computing diamonds. Its overall approach is illustrated by Example 1. That approach is to repeatedly identify an attribute value that cannot be in the diamond, and then (possibly not immediately) remove the attribute value and its slice. The identification of "bad" attribute values is done conservatively, in that they are known already to have a sum less than required ($\sigma$ is sum), or insufficient allocated cells ($\sigma$ is count). When the algorithm terminates, we are left with only attribute values that meet the condition in every slice: a diamond.

**Example 1.** *Suppose we seek a 4,10-carat diamond in Table 1 using Algorithm 1. On a first pass, we can delete the attribute values "Chicago" and "TV" because their respective slices have sums below 10 and 4. On a second pass, value "Berlin," "Game console" and "DVD" can be removed because the sums of their slices were reduced by the removal of the values "Chicago" and "TV." The algorithm then terminates.*

Algorithms based on this approach will always terminate, though they might sometimes return an empty cube. The correctness of our algorithm is guaranteed by the following result.

**Theorem 3.** *Algorithm 1 is correct, that is, it always returns the $k_1, k_2, \ldots, k_d$-carat diamond.*

*Proof.* Because the diamond is unique, we need only show that the result of the algorithm, the cube $A$, is a diamond. If the result is not the empty cube, then dimension $D_i$ has at least value $k_i$ per slice, and hence it has $k_i$ carats. We only need to show that the result of Algorithm 1 is maximal: there does not exist a larger $k_1, k_2, \ldots, k_d$-carat cube.

Suppose $A'$ is such a larger $k_1, k_2, \ldots, k_d$-carat cube. Because Algorithm 1 begins with the whole cube $C$, there must be a time when, for the first time, one of the attribute values of $C$ belonging to $A'$ but not $A$ is deleted. This attribute is not written to the output file because its corresponding slice of dimension `dim` had value less than $k_{\texttt{dim}}$. At the time of deletion, this attribute's slice cannot have obtained more cells after it had been deleted, so it still has value less than $k_{\texttt{dim}}$. Let $C'$ be the cube at the instant before the attribute is deleted, with all attribute values deleted so far. We see that $C'$ is larger than or equal to $A'$ and therefore, slices in $C'$ corresponding to attribute values of $A'$ along dimension `dim` must have more than $k_{\texttt{dim}}$ carats. Therefore, we have a contradiction and must conclude that $A'$ does not exist and that $A$ is maximal. □

**For simplicity of exposition, in the rest of the paper, we assume that the number of carats is the same for all dimensions.**

Our algorithm employs a preprocessing step that iterates over the input file creating $d$ hash tables that map attributes to their $\sigma$-values. When $\sigma = \text{COUNT}$, the $\sigma$-values for each dimension form a histogram, which might be precomputed in a DBMS.

These values can be updated quickly as long as $\sigma$ is linear: aggregators like SUM and COUNT are good candidates. If the cardinality of any of the dimensions is such that hash tables cannot be stored in main memory, then a file-based set of hash tables could be constructed. However, given a $d$-dimensional cube,

```
input: file inFile containing d−dimensional cube C, integer k > 0
output: the diamond data cube
// preprocessing scan computes σ values for each slice
foreach dimension i do
    Create hash table ht_i
    foreach attribute value v in dimension i do
        if σ( slice for value v of dimension i in C) ≥ k then
            ht_i(v) = σ( slice for value v of dimension i in C)
        end
    end
end
stable ← false
while ¬stable do
    Create new output file outFile // iterate main loop
    stable ← true
    foreach row r of inFile do
        (v_1, v_2, ..., v_d) ← r
        if v_i ∈ dom ht_i, for all 1 ≤ i ≤ d then
            write r to outFile
        else
            for j ∈ {1, ..., i−1, i+1, ..., d} do
                if v_j ∈ dom ht_j then
                    ht_j(v_j) = ht_j(v_j) − σ({r})
                    if ht_j(v_j) < k then
                        remove v_j from dom ht_j
                    end
                end
            end
            stable ← false
        end
    end
    if ¬stable then
        inFile ← outFile // prepare for another iteration
    end
end
return outFile
```

**Algorithm 1**: Diamond dicing for relationally stored cubes. Each iteration, less data is processed.

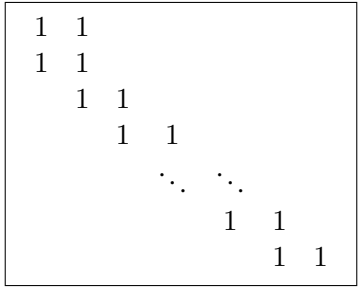

Figure 3: An $n \times n$ cube with $2n$ allocated cells (each indicated by a 1) and a 2-carat diamond in the upper left: it is a difficult case for an iterative algorithm.

there are only $\sum_{i=1}^{d} n_i$ slices and so the memory usage is $O(\sum_{i=1}^{d} n_i)$: for our tests, main memory hash tables suffice.

Algorithm 1 reads and writes the files sequentially from and to disk and does not require potentially expensive random access, making it a candidate for a data parallel implementation in the future.

Let $I$ be the number of iterations through the input file till convergence; ie no more deletions are done. Value $I$ is data dependent and (by Fig. 3) is $\Theta(\sum_i n_i)$ in the worst case. In practice, we do not expect $I$ to be nearly so large, and working with our largest "real world" data sets we never found $I$ to exceed 100.

Algorithm 1 runs in time $O(Id|C|)$; each attribute value is deleted at most once. In many cases, the input file decreases substantially in the first few iterations and those cubes will be processed faster than this bound suggests. The more carats we seek, the faster the file will decrease initially.

The speed of convergence of Algorithm 1 and indeed the size of an eventual diamond may depend on the data-distribution skew. Cell allocation in data cubes is very skewed and frequently follows Zipfian/-Pareto/zeta distributions [24]. Suppose the number of allocated cells $C_{\text{dim},i}$ in a given slice $i$ follows a zeta distribution: $P(C_{\text{dim},i} = j) \propto j^{-s}$ for $s > 1$. The parameter $s$ is indicative of the skew. We then have that $P(C_{\text{dim},i} < k_i) = \sum_{j=1}^{k_i-1} j^{-s} / \sum_{j=1}^{\infty} j^{-s} = P_{k_i,s}$. The expected number of slices marked for deletion after one pass of over all dimensions using $\sigma = \text{COUNT}$, prior to any slice deletion, is thus $\sum_{i=1}^{d} n_i P_{k_i,s}$. This quantity grows fast to $\sum_{i=1}^{d} n_i$ (all slices marked for deletion) as $s$ grows (see Fig. 4). For SUM-based diamonds, we not only have the skew of the cell allocation, but also the skew of the measures to accelerate convergence. In other words, we expect Algorithm 1 to converge quickly over real data sets, but more slowly over synthetic cubes generated using uniform distributions.

## 5.1 Finding the Largest Number of Carats

The determination of $\kappa(C)$, the largest value of $k$ for which $C$ has a non-trivial diamond, is a special case of the computation of the diamond-cube lattice (see Proposition 2). Identifying $\kappa(C)$ may help guide analysis. Two approaches have been identified:

1. Assume $\sigma = \text{COUNT}$. Set the parameter $k$ to 1 + the lower bound (provided by Proposition 6 or Theorem 2) and check whether there is a diamond with $k$ carats. Repeat, incrementing $k$, until an empty cube results. At each step, Proposition 2 says we can start from the cube from the previous iteration, rather than from $C$. When $\sigma$ is SUM, there are two additional complications. First, the value

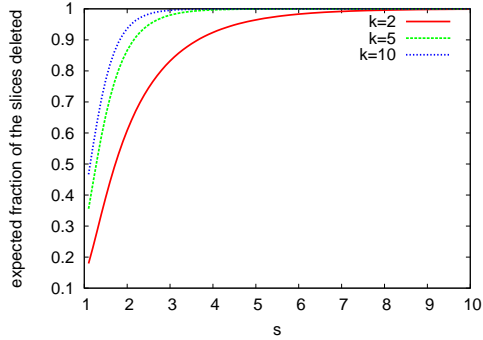

Figure 4: Expected fraction of slices marked for deletion after one pass under a zeta distribution for various values of the skew parameter $s$.

   of $k$ can grow large if measure values are large. Furthermore, if some measures are not integers, the result need not be an integer (hence we would compute $\lfloor \kappa(C) \rfloor$ by applying this method, and not $\kappa(C)$).

2. Assume $\sigma =$ COUNT. Observe that $\kappa(C)$ is in a finite interval. We have a lower bound from Proposition 6 or Theorem 2 and an upper bound $\prod_{i=1}^{d-1} n_i$ or $|C|$. (If this upper bound is unreasonably large, we can either use the number of cells in our current cube, or we could start with the lower bound and repeatedly double it.) Execute the diamond-dicing algorithm and set $k$ to a value determined by a binary search over its valid range. Every time the lower bound changes, we can make a copy of the resulting diamond. Thus, each time we test a new midpoint $k$, we can begin the computation from the copy (by Proposition 2). If $\sigma$ is SUM and measures are not integer values, it might be difficult to know when the binary search has converged exactly.

We believe the second approach is better. Let us compare one iteration of the first approach (which begins with a $k$-carat diamond and seeks a $k+1$-carat diamond) and a comparable iteration of the second approach (which begins with a $k$-carat diamond and seeks a $(k+k_{\text{upper}})/2$-carat diamond). Both will end up making at least one scan, and probably several more, through the $k$-carat diamond. Now, we experimentally observe that $k$ values that slightly exceed $\kappa(C)$ tend to lead to several times more scans through the cube than with other values of $k$. Our first approach will make only one such unsuccessful attempt, whereas the binary search would typically make several unsuccessful attempts while narrowing in on $\kappa(C)$. Nevertheless, we believe the fewer attempts will far outweigh this effect. We recommend binary search, given that it will find $\kappa(C)$ in O($\log \kappa(C)$) iterations.

If one is willing to accept an approximate answer for $\kappa(C)$ when aggregating with SUM, a similar approach can be used.

## 5.2 Diamond-Based Heuristic for DCLD

In Section 4.4, we noted that a diamond with the appropriate shape will not necessarily solve the DCLD problem. Nevertheless, when we examined many small random cubes, the solutions typically coincided. Therefore, we suggest diamond dicing as a heuristic for DCLD.

A heuristic for DCLD can start with a diamond and then refine its shape. Our heuristic first finds a diamond that is only somewhat too large, then removes slices until the desired shape is obtained. See Algorithm 2.

---

**input**: $d$-dimensional cube $C$, integers $p_1, p_2, \ldots p_d$
**output**: Cube with size $p_1 \times p_2 \times \ldots \times p_d$
`// Use binary search to find k`
Find max $k$ where the $k$-carat diamond $\Delta$ has shape $p'_1 \times p'_2 \times \ldots \times p'_d$, where $\forall i. p'_i \geq p_i$
**for** $i \leftarrow 1$ *to* $d$ **do**
    Sort slices of dimension $i$ of $\Delta$ by their $\sigma$ values
    Retain only the top $p_i$ slices and discard the remainder from $\Delta$
**end**
**return** $\Delta$

---

**Algorithm 2**: DCLD heuristic that starts from a diamond.

## 6 Experiments

We wish to show that diamonds can be computed efficiently. We also want to review experimentally some of the properties of diamonds including their density (count-based diamonds) and the range of values the carats may take in practice. Finally, we want to provide some evidence that diamond dicing can serve as the basis for a DCLD heuristic.

### 6.1 Data Sets

We experimented with diamond dicing on several different data sets, some of whose properties are laid out in Tables 2 and 5.

Cubes TW1, TW2 and TW3 were extracted from TWEED [12], which contains over 11,000 records of events related to internal terrorism in 18 countries in Western Europe between 1950 and 2004. Of the 52 dimensions in the TWEED data, 37 were measures since they decomposed the number of people killed/injured into all the affected groups. Cardinalities of the dimensions ranged from 3 to 284. Cube TW1 retained dimensions Country, Year, Action and Target with cardinalities of 16 × 53 × 11 ×11. For cubes TW2 and TW3 all dimensions not deemed measures were retained. Cubes TW2 and TW3 were rolled-up and stored

Table 2: Real data sets used in experiments

|  | TWEED | | | Netflix | | Census income | |
|---|---|---|---|---|---|---|---|
| cube | TW1 | TW2 | TW3 | NF1 | NF2 | C1 | C2 |
| dimensions | 4 | 15 | 15 | 3 | 3 | 28 | 28 |
| $\|C\|$ | 1957 | 4963 | 4963 | 100,478,158 | 100,478,158 | 196054 | 196054 |
| $\sum_{i=1}^{d} n_i$ | 88 | 674 | 674 | 500,137 | 500,137 | 533 | 533 |
| measure | count | count | killed | count | rating | stocks | wage |
| iters to converge | 6 | 10 | 3 | 19 | 40 | 6 | 4 |
| $\kappa$ | 38 | 37 | 85 | 1,004 | 3,483 | 99,999 | 9,999 |

in a MySQL database using the following query and the resulting tables were exported to comma separated
files. A similar process was followed for TW1. Table 3 lists the details of the TWEED data.

**INSERT INTO** tweed15( d1, d2, d3, d4, d5, d6, d7, d8,
    d31, d32, d33, d34, d50, d51, d52, d49 )
**SELECT** d1, d2, d3, d4, d5, d6, d7, d8,
    d31, d32, d33, d34, d50, d51, d52, **sum**( d9 )
**FROM** 'tweed'
**GROUP BY** ( d1, d2, d3, d4, d5, d6, d7, d8,
    d31, d32, d33, d34, d50, d51, d52 )

We also processed the Netflix data set [25], which has dimensions: MovieID × UserID × Date × Rating ($17766 \times 480189 \times 2182 \times 5$). Each row in the fact table has a distinct pair of values (MovieID, UserID). We extracted two 3-D cubes NF1 and NF2 both with about $10^8$ allocated cells using dimensions MovieID, UserID and Date. For NF2 we use Rating as the measure and the SUM aggregator, whereas NF1 uses the COUNT aggregator. The Netflix data set is the largest openly available movie-rating database ($\approx 2\,\text{GiB}$).

Our third real data set, Census-Income, comes from the UCI KDD Archive [17]. The cardinalities of the dimensions ranged from 2 to 91 and there were 199,523 records. We rolled-up the original 41 dimensions to 27 and used two measures, income from stocks(C1) and hourly wage(C2). The MySQL query used to generate cube C1 follows. Note that the dimension numbers map to those given in the census-income.names file [17]. Details are provided in table 4

**INSERT INTO** census_income_stocks( 'd0', 'd1', 'd2', 'd3', 'd4', 'd6',
    'd7', 'd8', 'd9', 'd10', 'd12', 'd13', 'd15', 'd21', 'd23',
    'd24', 'd25', 'd26', 'd27', 'd28', 'd29', 'd31', 'd32', 'd33',
    'd34', 'd35', 'd38', 'd18')
**SELECT** 'd0', 'd1', 'd2', 'd3', 'd4', 'd6', 'd7', 'd8', 'd9', 'd10',
    'd12', 'd13', 'd15', 'd21' 'd23', 'd24', 'd25', 'd26',
    'd27', 'd28', 'd29', 'd31', 'd32', 'd33', 'd34', 'd35', 'd38', **sum**( 'd18')
**FROM** census_income
**GROUP BY** 'd0', 'd1', 'd2', 'd3', 'd4', 'd6', 'd7', 'd8', 'd9',
    'd10', 'd12', 'd13', 'd15', 'd21', 'd23', 'd24', 'd25', 'd26', 'd27',
    'd28', 'd29', 'd31', 'd32', 'd33', 'd34', 'd35', 'd38';

We also generated synthetic data. As has already been stated, cell allocation in data cubes is skewed. We modelled this by generating values in each dimension that followed a power distribution. The values in dimension $i$ were generated as $\lfloor n_i u^{1/a} \rfloor$ where $u \in [0, 1]$ is a uniform distribution. For $a = 1$, this function generates uniformly distributed values. The dimensions are statistically independent. We picked the first 250,000 distinct facts. Since cubes S2A and S3A were generated with close to 250,000 distinct facts we decided to keep them all.

The cardinalities for all synthetic cubes are laid out in Table 6. All experiments on our synthetic data were done using the measure COUNT.

Table 3: Measures and dimensions of TWEED data. Shaded dimensions are those retained for TW1. All dimensions were retained for cubes TW2 and TW3 (with total people killed as its measure)

| | Dimension | Dimension cardinality |
|---|---|---|
| d1 | Day | 32 |
| d2 | Month | 13 |
| d3 | Year | 53 |
| d4 | Country | 16 |
| d5 | Type of agent | 3 |
| d6 | Acting group | 287 |
| d7 | Regional context of the agent | 34 |
| d8 | Type of action | 11 |
| d31 | State institution | 6 |
| d32 | Kind of action | 4 |
| d33 | Type of action by state | 7 |
| d34 | Group against which the state action is directed | 182 |
| d50 | Group's attitude towards state | 6 |
| d51 | Group's ideological character | 9 |
| d52 | Target of action | 11 |
| | Measure | |
| d49 | total people killed | |
| people from the acting group | military | police |
| civil servants | politicians | business executives |
| trade union leaders | clergy | other militants |
| civilians | | |
| total people injured | | |
| acting group | military | police |
| civil servants | politicians | business |
| trade union leaders | clergy | other militants |
| civilians | | |
| total people killed by state institution | | |
| group members | other people | |
| total people injured by state institution | | |
| group members | other people | |
| arrests | convictions | executions |
| total killed by non-state group at which the state directed an action | | |
| people from state institution | others | |
| total injured by non-state group | | |
| people from state institution | others | |

Table 4: Census Income data: dimensions and cardinality of dimensions. Shaded dimensions and measures retained for cubes C1 and C2. Dimension numbering maps to those described in the file *census-income.names* [17]

| | Dimension | Dimension cardinality |
|---|---|---|
| d0  | age | 91 |
| d1  | class of worker | 9 |
| d2  | industry code | 52 |
| d3  | occupation code | 47 |
| d4  | education | 17 |
| d6  | enrolled in education last week | 3 |
| d7  | marital status | 7 |
| d8  | major industry code | 24 |
| d9  | major occupation code | 15 |
| d10 | race | 5 |
| d12 | sex | 2 |
| d13 | member of a labour union | 3 |
| d15 | full or part time employment status | 8 |
| d21 | state of previous residence | 51 |
| d23 | detailed household summary in household | 8 |
| d24 | migration code - change in msa | 10 |
| d25 | migration code - change in region | 9 |
| d26 | migration code - moved within region | 10 |
| d27 | live in this house 1 year ago | 3 |
| d28 | migration previous residence in sunbelt | 4 |
| d29 | number of persons worked for employer | 7 |
| d31 | country of birth father | 43 |
| d32 | country of birth mother | 43 |
| d33 | country of birth self | 43 |
| d34 | citizenship | 5 |
| d35 | own business or self employed | 3 |
| d38 | weeks worked in year | 53 |
| d11 | hispanic origin | 10 |
| d14 | reason for unemployment | 6 |
| d19 | tax filer status | 6 |
| d20 | region of previous residence | 6 |
| d22 | detailed household and family status | 38 |
| ignored | instance weight | |
| d30 | family members under 18 | 5 |
| d36 | fill inc questionnaire for veteran's admin | 3 |
| d37 | veteran's benefits | 3 |
| d39 | year | 2 |
| ignored | classification bin | |
| | Measure | Cube |
| d18 | dividends from stocks | C1 |
| d5  | wage per hour | C2 |
| d16 | capital gains | |
| d17 | capital losses | |

Table 5: Synthetic data sets used in experiments

| cube | S1A | S1B | S1C | S2A | S2B | S2C | S3A | S3B | S3C |
|---|---|---|---|---|---|---|---|---|---|
| dimensions | 4 | 4 | 4 | 8 | 8 | 8 | 16 | 16 | 16 |
| skew factor | 0.02 | 0.2 | 1.0 | 0.02 | 0.2 | 1.0 | 0.02 | 0.2 | 1.0 |
| $\|C\|$ | 250k | 250k | 250k | 251k | 250k | 250k | 262k | 250k | 250k |
| $\sum_{i=1}^{d} n_i$ | 11,106 | 11,098 | 11,110 | 22,003 | 22,195 | 22,220 | 38,354 | 44,379 | 44,440 |
| iters to converge | 12 | 9 | 2 | 6 | 12 | 12 | 8 | 21 | 6 |
| $\kappa$ | 135 | 121 | 30 | 133 | 32 | 18 | 119 | 8 | 15 |

Table 6: Dimensional cardinalities for our synthetic data cubes

| Cube | Dimensional cardinalities |
|---|---|
| S1A | $6 \times 100 \times 1000 \times 10000$ |
| S1B | $2 \times 100 \times 1000 \times 9996$ |
| S1C | $10 \times 100 \times 1000 \times 10000$ |
| S2A | $10 \times 100 \times 1000 \times 9881 \times 10 \times 100 \times 1000 \times 9902$ |
| S2B | $10 \times 100 \times 1000 \times 9987 \times 10 \times 100 \times 1000 \times 9988$ |
| S2C | $10 \times 100 \times 1000 \times 10000 \times 10 \times 100 \times 1000 \times 10000$ |
| S3A | $10 \times 100 \times 1000 \times 8465 \times 10 \times 100 \times 1000 \times 8480$ $\times 10 \times 100 \times 1000 \times 8502 \times 10 \times 100 \times 1000 \times 8467$ |
| S3B | $10 \times 100 \times 1000 \times 9982 \times 10 \times 100 \times 1000 \times 9987$ $\times 10 \times 100 \times 1000 \times 9988 \times 10 \times 100 \times 1000 \times 9982$ |
| S3C | $10 \times 100 \times 1000 \times 10000 \times 10 \times 100 \times 1000 \times 10000$ $\times 10 \times 100 \times 1000 \times 10000 \times 10 \times 100 \times 1000 \times 10000$ |

All experiments were carried out on a Linux-based (Ubuntu 7.04) dual-processor machine with Intel Xeon (single core) 2.8 GHz processors with 2 GiB RAM. It had one disk, a Seagate Cheetah ST373453LC (SCSI 320, 15 kRPM, 68 GiB), formatted to the ext3 filesystem. Our implementation was done with Sun's SDK 1.6.0 and to handle the large hash tables generated when processing Netflix, we set the maximum heap size for the JVM to 2 GiB.

## 6.2 Iterations to Convergence

Algorithm 1 required 19 iterations and an average of 35 minutes to compute the 1004-carat $\kappa$-diamond for NF1. However it took 50 iterations and an average of 60 minutes to determine that there was no 1005-carat diamond. The preprocessing time for NF1 was 22 minutes. For a comparison, sorting the Netflix comma-separated data file took 29 minutes. Times were averaged over 10 runs. Fig. 5 shows the number of cells present in the diamond after each iteration for 1004–1006 carats. The curve for 1006 reaches zero first, followed by that for 1005. Since $\kappa(\text{NF1}) = 1004$, that curve stabilizes at a nonzero value. We see a similar result for TW2 in Fig. 6 where $\kappa$ is 37. It takes longer to reach a critical point when $k$ only slightly exceeds $\kappa$.

As stated in Section 5, the number of iterations required until convergence for all our real and synthetic cubes was far fewer than the upper bound, e.g. cube S2B: 2,195 (upper bound) and 12 (actual). We had expected to see the uniformly distributed data taking longer to converge than the skewed data. This was not the case. It may be that a clearer difference would be apparent on larger synthetic data sets. This will be investigated in future experiments.

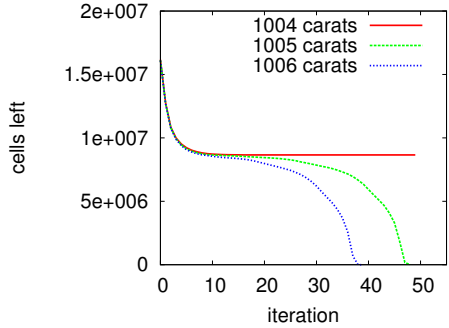

Figure 5: Cells remaining after each iteration of Algorithm 1 on NF1, computing a 1004-, 1005- and 1006-carat diamonds.

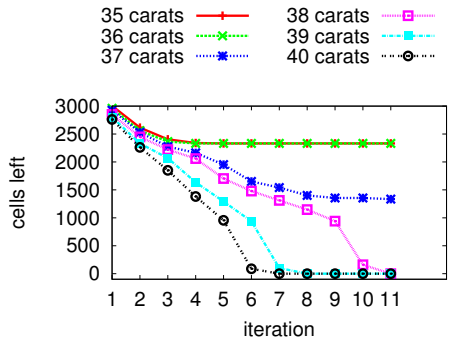

Figure 6: Cells remaining after each iteration, TW2

## 6.3 Largest Carats

According to Proposition 6, COUNT-$\kappa$(NF1) $\geq$ 197. Experimentally, we determined that it was 1004. By the definition of the carat, it means we can extract a subset of the Netflix data set where each user entered at least 1004 ratings on movies rated at least 1004 times by these same users during days where there were at least 1004 ratings by these same users on these same movies. The 1004-carat diamond had dimensions $3082 \times 6833 \times 1351$ and 8,654,370 cells, for a density of about $3 \times 10^{-4}$ or two orders of magnitude denser than the original cube. The presence of such a large diamond was surprising to us. We believe nothing similar has been observed about the Netflix data set before [5].

Comparing the two methods in Section 5.1, we see that sequential search would try 809 values of $k$ before identifying $\kappa$. However, binary search would try 14 values of $k$ (although 3 are between 1005 and 1010, where perhaps double or triple the normal number of iterations are required). To test the time difference for the two methods, we used cube TW1. We executed a binary search, repeatedly doubling our lower bound to obtain the upper limit, and thus until we established the range where $\kappa$ must exist. Whenever we exceeded $\kappa$, a copy of the original data was used for the next step. Even with this copying step and the unnecessary recomputation from the original data, the time for binary search averaged only 2.75 seconds. Whereas a sequential search, that started with the lower bound and increased $k$ by one, averaged 9.854 seconds over ten runs.

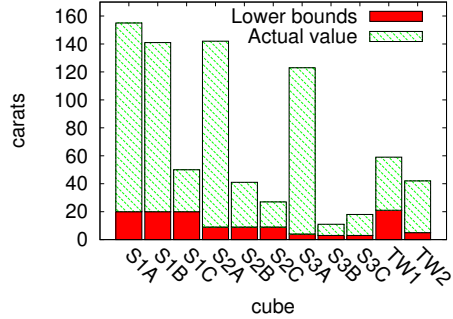

Figure 7: Comparison between estimated $\kappa$, based on the lower bounds from Proposition 6, and number of (COUNT-based) carats found.

Fig. 7 shows our lower bounds on $\kappa$, given the dimensions and numbers of allocated cells in each cube, compared with their actual $\kappa$ values. The plot indicates that our lower bounds are further away from actual values as the skew of the cube increases for the synthetic cubes. Also, we are further away from $\kappa$ for TW2, a cube with 15 dimensions, than for TW1. For uniformly-distributed cubes S1C, S2C and S3C there was no real difference in density between the cube and its diamond. However, all other diamonds experienced an increase of between 5 and 9 orders of magnitude.

Diamonds found in C1, C2, NF2 and TW3 captured 0.35%, 0.09%, 66.8% and 0.6% of the overall sum for each cube respectively. The very small fraction captured by the diamond for TW3 can be explained by the fact that $\kappa$(TW3) is based on a diamond that has only one cell, a bombing in Bologna in 1980 that killed 85 people. Similarly, the diamond for C2 also comprised a single cell.

## 6.4 Effectiveness of DCLD Heuristic

To test the effectiveness of our diamond-based DCLD heuristic (Subsection 5.2), we used cube TW1 and set the parameter $p$ to 5. We were able to establish quickly that the 38-carat diamond was the closest to satisfying this constraint. It had density of 0.169 and cardinalities of $15 \times 7 \times 5 \times 8$ for the attribute values; year, country, action and target. The solution we generated to this DCLD ($p = 5$) problem had exactly 5 attribute values per dimension and density of 0.286.

Since the DCLD problem is NP-complete, determining the quality of the heuristic poses difficulties. We are not aware of any known approximation algorithms and it seems difficult to formulate a suitably fast exact solution by, for instance, branch and bound. Therefore, we also implemented a second computationally expensive heuristic, in hope of finding a high-quality solution with which to compare our diamond-based heuristic. This heuristic is based on local search from an intuitively reasonable starting state. (A greedy steepest-descent approach is used; states are $(\langle A_1, A_2, \ldots, A_d \rangle$, where $|A_i| = p_i$, and the local neighbourhood of such a state is $\langle A'_1, A'_2, \ldots, A'_d \rangle$, where $A_i = A'_i$ except for one value of $i$, where $|A_i \cap A'_i| = p_i - 1$.

The starting state consists of the most frequent $p_i$ values from each dimension $i$. Our implemention actually requires the $i^{th}$ local move be chosen along dimension $i \bmod d$, although if no such move brings improvement, no move is made.)

---

**input**: $d$-dimensional cube $C$, integers $p_1, p_2, \ldots p_d$
**output**: Cube with size $p_1 \times p_2 \times \ldots \times p_d$
**foreach** *dimension $i$* **do**
    Sort slices of dimension $i$ of $\Delta$ by their $\sigma$ values
    Retain only the top $p_i$ slices and discard the remainder from $\Delta$
**end**
**repeat**
    **for** $i \leftarrow 1$ *to* $d$ **do**
        // We find the best swap in dimension i
        bestAlternative $\leftarrow \sigma(\Delta)$
        **foreach** *value $v$ of dimension $i$ that has been retained in $\Delta$* **do**
            **foreach** *value $w$ from dimension $i$ in $C$, but where $w$ is not in $\Delta$* **do**
                Form $\Delta'$ by temporarily adding slice $w$ and removing slice $v$ from $\Delta$
                **if** $\sigma(\Delta') >$ bestAlternative **then**
                      $(\text{rem}, \text{add}) \leftarrow (v, w)$; bestAlternative $\leftarrow \sigma(\Delta')$
                **end**
            **end**
        **end**
        **if** bestAlternative $> \sigma(\Delta)$ **then**
            Modify $\Delta$ by removing slice rem and adding slice add
        **end**
    **end**
**until** $\Delta$ *was not modified by any $i$*
**return** $\Delta$

**Algorithm 3**: Expensive DCLD heuristic.

---

The density reported by Algorithm 3 was 0.283, a similar outcome, but at the expense of more work. Our diamond-based heuristic, starting with the 38-carat diamond, required a total of 15 deletes. Whereas our expensive comparision heuristic, starting with its $5 \times 5 \times 5 \times 5$ subcube, required 1420 inserts/deletes. Our diamond heuristic might indeed be a useful starting point for a solution to the DCLD problem.

### 6.5 Robustness against randomly missing data

We experimented with cube TW1 to determine whether diamond dicing appears robust against random noise that models the data warehouse problem [31] of missing data. Existing data points had an independent probability $p_{\text{missing}}$ of being omitted from the data set, and we show $p_{\text{missing}}$ versus $\kappa(\text{TW1})$ for 30 tests each with $p_{\text{missing}}$ values between 1% and 5%. Results are shown as in Table 7. Our answers were rarely more than 8% different, even with 5% missing data.

Table 7: Robustness of $\kappa(\mathsf{TW1})$ under various amount of randomly missing data: for each probability, 30 trials were made. Each column is a histogram of the observed values of $\kappa(\mathsf{TW1})$.

| $\kappa(\mathsf{TW1})$ | Prob. of cell's deallocation | | | | |
|---|---|---|---|---|---|
| | 1% | 2% | 3% | 4% | 5% |
| 38 | 19 | 12 | 3 | 2 | |
| 37 | 10 | 17 | 17 | 10 | 4 |
| 36 | 1 | 1 | 10 | 16 | 18 |
| 35 | | | | 2 | 7 |
| 34 | | | | | 1 |

# 7 Conclusion and Future Work

We introduced the diamond dice, a new OLAP operator that dices on all dimensions simultaneously. This new operation represents a multidimensional generalization of the iceberg query and can be used by analysts to discover sets of attribute values jointly satisfying multidimensional constraints.

We have shown that the problem is tractable. We were able to process the 2 GiB Netflix data with 500,000 distinct attribute values and 100 million cells in about 35 minutes, excluding preprocessing. As expected from the theory, real-world data sets have a fast convergence using Algorithm 1: the first few iterations quickly prune most of the false candidates. We have identified potential strategies to improve the performance further. First, we might selectively materialize elements of the diamond-cube lattice (see Proposition 2). The computation of selected components of the diamond-cube lattice also opens up several optimization opportunities. Second, we believe we can use ideas from the implementation of ITERATIVE PRUNING proposed by Kumar et al. [18]. Third, Algorithm 1 is suitable for parallelization [11]. Also, our current implementation uses only Java's standard libraries and treats all attribute values as strings. We believe optimizations can be made by the preprocessing step that will greatly reduce overall running time.

We presented theoretical and empirical evidence that a non-trivial, single, dense chunk can be discovered using the diamond dice and that it provides a sensible heuristic for solving the DENSEST CUBE WITH LIMITED DIMENSIONS. The diamonds are typically much denser than the original cube. Over moderate cubes, we saw an increase of the density by one order of magnitude, whereas for a large cube (Netflix) we saw an increase by two orders of magnitude and more dramatic increases for the synthetic cubes. Even though Lemma 2 states that diamonds do not necessarily have optimal density given their shape, informal experiments suggest that they do with high probability. This may indicate that we can bound the sub-optimality, at least in the average case; further study is needed.

We have shown that sum-based diamonds are no harder to compute than count-based diamonds and we plan to continue working towards an efficient solution for the HEAVIEST CUBE WITH LIMITED DIMENSIONS (HCLD).

# References

[1] C. Anderson. *The long tail*. Hyperion, 2006.

[2] K. Aouiche, D. Lemire, and R. Godin. Collaborative OLAP with tag clouds: Web 2.0 OLAP formalism and experimental evaluation. In *WEBIST'08*, 2008.

[3] B. Babcock, S. Chaudhuri, and G. Das. Dynamic sample selection for approximate query processing. In *SIGMOD'03*, pages 539–550, 2003.

[4] R. Ben Messaoud, O. Boussaid, and S. Loudcher Rabaséda. Efficient multidimensional data representations based on multiple correspondence analysis. In *KDD'06*, pages 662–667, 2006.

[5] J. Bennett and S. Lanning. The Netflix prize. In *KDD Cup and Workshop 2007*, 2007.

[6] S. Börzsönyi, D. Kossmann, and K. Stocker. The skyline operator. In *ICDE '01*, pages 421–430. IEEE Computer Society, 2001.

[7] M. J. Carey and D. Kossmann. On saying "enough already!" in SQL. In *SIGMOD'97*, pages 219–230, 1997.

[8] G. Cormode, F. Korn, S. Muthukrishnan, and D. Srivastava. Diamond in the rough: finding hierarchical heavy hitters in multi-dimensional data. In *SIGMOD '04*, pages 155–166, New York, NY, USA, 2004. ACM Press.

[9] G. Cormode and S. Muthukrishnan. What's hot and what's not: tracking most frequent items dynamically. *ACM Trans. Database Syst.*, 30(1):249–278, 2005.

[10] M. Dawande, P. Keskinocak, J. M. Swaminathan, and S. Tayur. On bipartite and multipartite clique problems. *Journal of Algorithms*, 41(2):388–403, November 2001.

[11] F. B. Dehne, T. B. Eavis, and A. B. Rau-Chaplin. The cgmCUBE project: Optimizing parallel data cube generation for ROLAP. *Distributed and Parallel Databases*, 19(1):29–62, 2006.

[12] J. O. Engene. Five decades of terrorism in Europe: The TWEED dataset. *Journal of Peace Research*, 44(1):109–121, 2007.

[13] M. Fang, N. Shivakumar, H. Garcia-Molina, R. Motwani, and J. D. Ullman. Computing iceberg queries efficiently. In *VLDB'98*, pages 299–310, 1998.

[14] V. Ganti, M. L. Lee, and R. Ramakrishnan. ICICLES: Self-tuning samples for approximate query answering. In *VLDB'00*, pages 176–187, 2000.

[15] R. Godin, R. Missaoui, and H. Alaoui. Incremental concept formation algorithms based on Galois (concept) lattices. *Computational Intelligence*, 11:246–267, 1995.

[16] J. Gray, A. Bosworth, A. Layman, and H. Pirahesh. Data cube: A relational aggregation operator generalizing group-by, cross-tab, and sub-total. In *ICDE '96*, pages 152–159, 1996.

[17] S. Hettich and S. D. Bay. The UCI KDD archive. http://kdd.ics.uci.edu, 2000. last checked April 28, 2008.

[18] R. Kumar, P. Raghavan, S. Rajagopalan, and A. Tomkins. Trawling the web for emerging cyber-communities. In *WWW '99*, pages 1481–1493, New York, NY, USA, 1999. Elsevier North-Holland, Inc.

[19] D. Lemire and O. Kaser. Hierarchical bin buffering: Online local moments for dynamic external memory arrays. *ACM Trans. Algorithms*, 4(1):1–31, 2008.

[20] C. Li, B. C. Ooi, A. K. H. Tung, and S. Wang. DADA: a data cube for dominant relationship analysis. In *SIGMOD'06*, pages 659–670, 2006.

[21] Z. X. Loh, T. W. Ling, C. H. Ang, and S. Y. Lee. Adaptive method for range top-k queries in OLAP data cubes. In *DEXA'02*, pages 648–657, 2002.

[22] Z. X. Loh, T. W. Ling, C. H. Ang, and S. Y. Lee. Analysis of pre-computed partition top method for range top-k queries in OLAP data cubes. In *CIKM'02*, pages 60–67, 2002.

[23] M. D. Morse, J. M. Patel, and H. V. Jagadish. Efficient skyline computation over low-cardinality domains. In *VLDB*, pages 267–278, 2007.

[24] T. P. E. Nadeau and T. J. E. Teorey. A Pareto model for OLAP view size estimation. *Information Systems Frontiers*, 5(2):137–147, 2003.

[25] Netflix, Inc. Nexflix prize. http://www.netflixprize.com, 2007. last checked April 28, 2008.

[26] R. Peeters. The maximum-edge biclique problem is NP-complete. Research Memorandum 789, Faculty of Economics and Business Administration, Tilberg University, 2000.

[27] J. Pei, M. Cho, and D. Cheung. Cross table cubing: Mining iceberg cubes from data warehouses. In *SDM'05*, 2005.

[28] R. G. Pensa and J. Boulicaut. Fault tolerant formal concept analysis. In *AI*IA 2005*, volume 3673 of *LNAI*, pages 212–233. Springer-Verlag, 2005.

[29] D. N. Politis, J. P. Romano, and M. Wolf. *Subsampling*. Springer, 1999.

[30] P. K. Reddy and M. Kitsuregawa. An approach to relate the web communities through bipartite graphs. In *WISE'01*, pages 302–310, 2001.

[31] E. Thomson. *OLAP Solutions: Building Multidimensional Information Systems*. Wiley, second edition, 2002.

[32] H. Webb. Properties and applications of diamond cubes. In *ICSOFT 2007 – Doctoral Consortium*, 2007.

[33] D. Xin, J. Han, X. Li, and B. W. Wah. Star-cubing: Computing iceberg cubes by top-down and bottom-up integration. In *VLDB*, pages 476–487, 2003.

[34] K. Yang. Information retrieval on the web. *Annual Review of Information Science and Technology*, 39:33–81, 2005.

[35] M. L. Yiu and N. Mamoulis. Efficient processing of top-k dominating queries on multi-dimensional data. In *VLDB'07*, pages 483–494, 2007.